\documentclass[a4paper,fleqn]{cas-dc}

\usepackage[numbers,sort&compress]{natbib}

\usepackage{graphicx}
\usepackage{siunitx}
\usepackage{bm}
\usepackage{lineno}
\usepackage{verbatim}

\usepackage[linesnumbered,ruled,vlined]{algorithm2e}
\usepackage{geometry}
\usepackage{enumitem}

\usepackage{booktabs}
\usepackage{caption}
\usepackage{multirow}

\usepackage{hyperref}
\hypersetup{
    colorlinks=true,
    citecolor=blue,
    linkcolor=blue,
    urlcolor=blue,
    filecolor=blue
}

\usepackage{xcolor}
\usepackage{float}
\usepackage{csquotes}
\usepackage{soul}
\usepackage{makecell}

\begin{document}
\let\WriteBookmarks\relax
\def\floatpagepagefraction{1}
\def\textpagefraction{.001}
\shorttitle{U-Net for short-time muon tomography}
\shortauthors{H. Wang et~al.}

\title [mode = title]{U-Net Based Image Enhancement for Short-time Muon Scattering Tomography}

\author[1]{Haochen Wang}
\fnmark[1]
\credit{Writing – original draft, Visualization, Software, Methodology}
\affiliation[1]{
    organization={School of Physics, Hefei University of Technology},
    city={Hefei},
    postcode={230601},
    country={China}
}

% 第二作者：Pei Yu（共同一作）
\author[2,3]{Pei Yu}
\fnmark[1]
\credit{Writing – original draft, Visualization, Software, Methodology}
\affiliation[2]{
    organization={Advanced Energy Science and Technology Guangdong Laboratory},
    city={Huizhou},
    postcode={516000},
    country={China}
}
\affiliation[3]{
    organization={Institute of Modern Physics, Chinese Academy of Sciences},
    city={Lanzhou},
    postcode={730000},
    country={China}
}

% 第三作者：Liangwen Chen（通讯作者）
\author[2,3,4,5]{Liangwen Chen}
\cormark[1]
\credit{Writing – review \& editing, Writing – original draft, Supervision, Software, Methodology}
\affiliation[4]{
    organization={School of Nuclear Science and Technology, University of Chinese Academy of Sciences},
    city={Beijing},
    postcode={100049},
    country={China}
}
\affiliation[5]{
    organization={State Key Laboratory of Heavy Ion Science and Technology, Institute of Modern Physics, Chinese Academy of Sciences},
    city={Lanzhou},
    postcode={730000},
    country={China}
}

% 第四作者：Weibo He（通讯作者）
\author[6]{Weibo He}
\cormark[2]
\credit{Writing – review \& editing, Writing – original draft, Supervision, Methodology}
\affiliation[6]{
    organization={Institute of Materials, China Academy of Engineering Physics},
    city={Jiangyou},
    postcode={621907},
    country={China}
}

% 第五作者：Yu Zhang（通讯作者）
\author[1]{Yu Zhang}[orcid=0000-0001-9415-8252]
\cormark[3]
\credit{Writing – review \& editing, Writing – original draft, Supervision, Methodology}

% 第六作者：Yuhong Yu
\author[2,3,4,5]{Yuhong Yu}
\credit{Writing – review \& editing}

% 第七作者：Xueheng Zhang
\author[2,3,4,5]{Xueheng Zhang}
\credit{Writing – review \& editing}

% 第八作者：Lei Yang
\author[2,3,4,5]{Lei Yang}
\credit{Writing – review \& editing}

% 第九作者：Zhiyu Sun
\author[2,3,4,5]{Zhiyu Sun}
\credit{Writing – review \& editing}

%%% 脚注说明 %%%
\fntext[fn1]{These authors contributed equally to this work.}

%%% 通讯作者说明 %%%
\cortext[cor1]{Corresponding author: chenlw@impcas.ac.cn}
\cortext[cor2]{Corresponding author: njuyyf@163.com}
\cortext[cor3]{Corresponding author: dayu@hfut.edu.cn}

\begin{abstract}
Muon Scattering Tomography (MST) is a promising non-invasive inspection technique, yet the practical application of short-time MST is hindered by poor image quality due to limited muon flux. To address this limitation, we propose a U-Net-based framework trained on Point of Closest Approach (PoCA) images reconstructed with simulation MST data to enhance image quality. When applied to experimental MST data, the framework significantly improves image quality, increasing the Structural Similarity Index Measure (SSIM) from 0.7232 to 0.9699 and decreasing the Learned Perceptual Image Patch Similarity (LPIPS) from 0.3604 to 0.0270. These results demonstrate that our method can effectively enhance low-statistics MST images, thereby paving the way for the practical deployment of short-time MST.
\end{abstract}


\maketitle

\section{\label{sec:Introduction}Introduction}
Cosmic ray muons are produced from the decay of mesons generated when natural cosmic rays interact with atomic nuclei in the Earth's atmosphere, with a flux of about \SI{1}{min^{-1}cm^{-2}} \cite{Shukla2018, Neddermeyer1938, Su2021}. The natural cosmic ray muons are characterized by their strong penetration capability, absence of radiation damage, and high sensitivity to materials with high atomic numbers (high Z), such as nuclear materials, making them a non-destructive imaging technique with unique advantages. In 1955, E. P. George first applied cosmic ray muons to estimate the overlay above underground tunnels in Australia \cite{George:1955bzp}. In 1970, Alvarez used muon technology to detect hidden chambers in the pyramids \cite{Alvarez1970}. In 2003, based on the principle of Multiple Coulomb Scattering from the interaction between muons and matter, the Los Alamos National Laboratory (LANL) first proposed the Muon Tomography algorithm named Point of Closest Approach (PoCA),  and applied it to the inspection of border-crossing containers \cite{Borozdin2003}. Subsequently, a series of Muon Scattering Tomography (MST) algorithms have emerged, including the statistical iterative reconstruction algorithm MLSD based on scattering angle and displacement \cite{Schultz2007}, the MAP algorithm based on maximizing the Bayesian posterior probability \cite{GuobaoWang2009}, and algorithms based on the estimation of the Most Probable Trajectory \cite{Yi2014}. Since then, muon technology has been widely applied in fields such as border security \cite{Barnes2023}, studies of buildings and underground tunnels \cite{Thompson2020, Guardincerri2017, Liu2023}, and nuclear safety \cite{Lefevre2025, Bae2024}.

Due to intrinsic noise, such as environmental background noise and electronic noise from the detector measurement process, as well as scattering approximations in traditional imaging algorithms like PoCA, the resulting images are often of poor quality. Concurrently, the low-flux nature of cosmic ray muons significantly increases the time cost required to obtain sufficient muon event statistics for clear structural imaging. These are the key issues constraining the deployment and popularization of muon imaging technology, especially in the field of material monitoring \cite{Poulson2017, Braunroth2021}. Deep learning technology, particularly Convolutional Neural Networks (CNNs), has achieved remarkable success in the field of image processing and optimization. In 2015, Ronneberger proposed the U-Net \cite{https://doi.org/10.48550/arxiv.1505.04597}, which achieved significant accomplishments in the field of image segmentation. Subsequently, U-Net and its variants have achieved a series of successes in fields such as medical PET and CT imaging optimization. Kaviani used a transformer U-Net to optimize low-dose PET images while preserving important features \cite{Kaviani2023}. Wang proposed a sparse-view CT reconstruction method based on Filtered Swin Transformer U-Net \cite{Wang2025}.

However, related applications in the field of muon tomography have been less studied. We introduce a tailored U-Net to perform the dual optimization task of denoising and muon signal enhancement, thereby addressing bottlenecks of MST. Based on the images reconstructed from the simulated MST data, this study proposes a training framework with the U-Net model to enhance the image quality in short-time MST. With a high-quality PoCA image reconstructed from simulated muon data as the label, we first prepare a diverse dataset composed of PoCA images reconstructed from simulation MST data, with a wide range of event levels and detector resolutions. Subsequently, we propose a novel cross-domain style injection method, termed \enquote{Stamping}, for dataset augmentation. This approach augments our reconstructed PoCA images of simulated MST data by embedding the unique noise characteristics of PoCA images of experimental MST data, creating a highly effective hybrid training dataset that could further generalize the model's enhancement ability. Through a series of studies and image quality assessments, we identify the optimal strategy for configuring the dataset, and also validate the effectiveness of the proposed method with PoCA images of experimental MST data. The trained U-Net acts as an efficient post-processor, showing exceptional performance in significantly enhancing the quality of the images reconstructed from low-statistics experimental data. Our method significantly reduces the time cost and the requirements for detector quality of muon imaging in the field of structural monitoring, which is of great importance for the popularization of muon imaging algorithms.

\section{\label{sec:method}Method}
In the method section, we introduce the U-Net architecture employed in this work. Next, we elaborate on the acquisition of muon trajectory data for subsequent PoCA reconstruction, covering both the real experimental data for images to be enhanced and the simulated data for images as training datasets. We present an innovative dataset augmentation strategy, termed \enquote{Stamping}, which is designed to generate a hybrid dataset for sim-to-real domain adaptation training. Finally, we define the set of metrics used for the quantitative evaluation of the image quality.

\begin{table}[!htb]
	\centering
	\caption{Parameter setting in training stage.}
	\label{tab:parameters}
	
	\renewcommand{\arraystretch}{1.4}
	\begin{tabular}{>{\raggedright\arraybackslash}p{2.1cm}
                    >{\raggedright\arraybackslash}p{1.0cm}
                    >{\raggedright\arraybackslash}p{4.0cm}}
	
		\toprule[1pt]
		\textbf{Parameter} & \textbf{Value} & \textbf{Description} \\
		\midrule
            in/out\_dim & $300^2$ & Input and output images are single-channel images \\
            dbcov\_kernel & 3 & The kernel size in DoubleConv \\
            padding & 1 & Before DoubleConv to ensure the space dimensions unchanged \\
            activation & ReLU & The activation function used after each convolution in DoubleConv \\
            batchnorm & True & BatchNorm after each convolution in DoubleConv \\
            maxpool\_kernel & 2 & The kernel size of MaxPool2d used for downsampling \\
            batch\_size & 4 & Choose appropriately based on the video memory\\
            optimizer & Adam & Import Adam from \textit{torch} as the optimizer\\
            $\alpha$ & 0.7 & The weight of LPIPS loss in the joint loss function\\
            lr & 1e-4 & The learning rate in the training process\\
            epoch & 300 & The number of epochs in the training stage \\
		\bottomrule[1pt]
	\end{tabular}
\end{table}

\subsection{\label{ssec:unet}U-Net model and training}
Our model is a tailored U-Net with a classic encoder-decoder structure, where skip connections fuse low-level spatial details with high-level semantic features. The network is constructed from Double Convolution Blocks and enhanced with two key modifications for improved performance: Batch Normalization is added after each convolution to stabilize and accelerate training, and consistent padding is used to ensure the output dimensions match the input.  The encoder progressively downsamples the input through four stages, while the decoder symmetrically upsamples the feature maps. After the final upsampling stage, a terminal $1\times1$ convolutional layer maps the 64-channel feature map to a single-channel output, yielding the final optimized muon tomography image.

\begin{figure*}[!htb]
\centering
\includegraphics[width=\textwidth]{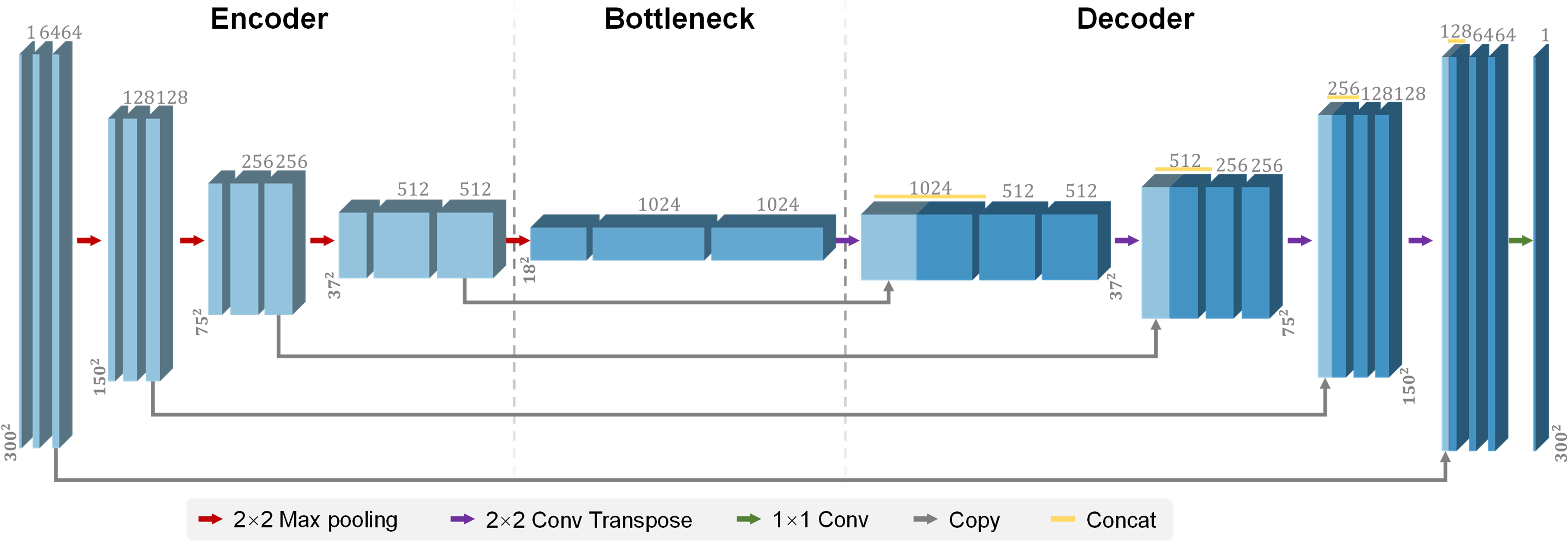}
\caption{The structure of the U-Net in our studies, the encoder, bottleneck layer, and decoder are displayed in different blue depths}
\label{fig:unet}
\end{figure*}

We employ a joint loss function that combines the $L_1$ loss and the image quality assessment (IQA) \cite{5596999, Yu_2016, ZhouWang2004, Zhang2018} term to train the model:
\begin{equation}
    \mathcal{L} = (1-\alpha) {\mathcal{L}_{L_1}}+\alpha {\mathcal{L}_{IQA}},
\label{eq:loss_function}
\end{equation}
where
\begin{equation}
    {\mathcal{L}_{L_1}} = \frac{1}{N}\sum_{i=1}^N |X_i - Y_i|.
\label{eq:loss_function_term}
\end{equation}
The $L_1$ loss enforces pixel-wise accuracy, while the IQA loss term calculates the difference between the output image and the target image using the specified method. The calculation method of the IQA loss term is similar to the metrics detailed in Section. \ref{ssec:quality}. This joint loss results in a better training performance compared with the traditional single loss term. The model's performance based on different IQA loss training will be discussed in Section. \ref{ssec:loss}. The specific network structure and parameter settings are detailed in Fig. \ref{fig:unet} and Table. \ref{tab:parameters}. The model was implemented using the \textit{PyTorch} framework. All training and inference procedures were conducted on a workstation equipped with an \textit{Intel Core i9-14900K CPU} and a \textit{NVIDIA GeForce RTX 3050 GPU} with 6 GB of VRAM.

\subsection{\label{ssec:dataset}Datasets}

\subsubsection{\label{sssec:det}Experimental data}
The experimental muon data were supported by the Muon Imaging Group at the University of Science and Technology of China \cite{Wang2022, Feng2021}, using a muon imaging system with eight layers of micromegas detectors. The upper group of four detectors was placed $\SI{100}{mm}$ above the tungsten block to measure the incoming muon tracks, and the lower group of four detectors was placed $\SI{64.3}{mm}$ below the block to measure the outgoing muon tracks. The gap between two detectors inside each group is $\SI{52.5}{mm}$. The imaging target, a C-shaped tungsten block, is centered within this gap. The block has outer dimensions of $\SI{8}{mm}\times\SI{8}{mm}\times\SI{4}{mm}$ and features a central void with $\SI{6}{mm}\times\SI{4}{mm}\times\SI{4}{mm}$. The effective area of each detector is $\SI{150}{mm}\times\SI{150}{mm}$, and the spatial resolution is about $\SI{100}{\micro\meter}$. The MST experiment lasted for 24 hours, during which 18,417 valid muon events were detected. We divided the collected muon data into 11 groups based on different random seeds, each of which contained 10,000 muon events. 11 independent PoCA images were reconstructed from these groups of data, named \textbf{Image-1} to \textbf{11}, for subsequent dataset augmentation, image enhancement, and method generalization tests.

\subsubsection{\label{sssec:sim}Simulated data}

\begin{figure}[!htb]
\includegraphics[width=\linewidth]{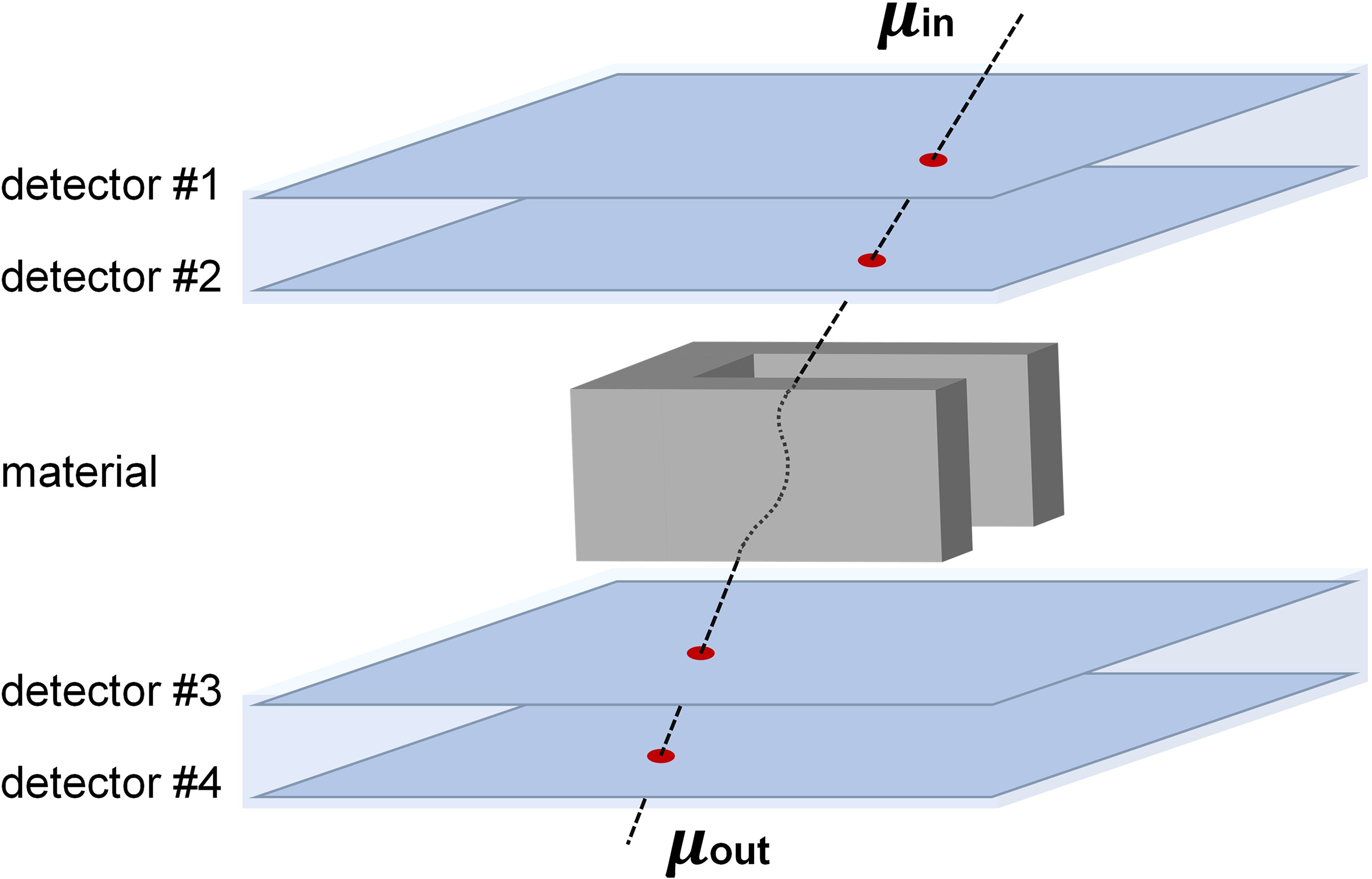}
\caption{Schematic of the Geant4 simulation setup.}
\label{fig:simulation}
\end{figure}

The training of the U-Net requires a sufficient number of diverse samples to achieve good performance. Unlike the medical imaging field, which has a large number of patient sample images available for training \cite{Chen2017, https://doi.org/10.48550/arxiv.1807.10165}, obtaining a large amount of muon data for generating images in the experimental scenario is often a very time-consuming process. To construct a large and diverse training dataset, a Monte Carlo simulation was developed using the Geant4 toolkit in conjunction with the Cosmic-Ray Shower Library (CRY) \cite{4437209}. As illustrated in Fig. \ref{fig:simulation}, the muon tomography setup consists of four position-sensitive detector planes, each with an active area of $\SI{150}{mm}\times\SI{150}{mm}$. The detectors are arranged in two groups (top: \#1, \#2; bottom: \#3, \#4) placed above and below the imaging volume. The spacing between two detectors in one group is $\SI{55}{mm}$, while the distance between the two groups (between detectors \#2 and \#3) is $\SI{135}{mm}$. The geometric structure and material of the imaging target strictly comply with the C-shaped structure used in the detector experiment.

An initial of 600,000 valid muon events was simulated for the generation of image samples corresponding to various muon event levels. The specific sampling strategy of the dataset is detailed in the Section. \ref{ssec:sampling}. Furthermore, to emulate the finite spatial resolution of physical detectors, we add 2D isotropic Gaussian noise $\mathcal{N}(0, \sigma^{2})$ at X and Y directions of the recorded hit positions on all four detectors. We simulated 10 distinct levels of spatial resolution by varying the standard deviation $\sigma$ of this noise from $\SI{0.1}{mm}$ to $\SI{1.0}{mm}$ in $\SI{0.1}{mm}$ increments. This strategy expands the dataset by an order of magnitude (number of event levels × 10 resolution levels), thereby increasing the diversity of the dataset and facilitating a more robust training process. Using these simulated data, PoCA images will be reconstructed and augmented, thereby forming the final training dataset. On the other hand, we selected all 600,000 muon events without position noise to generate a high-quality PoCA image, which serves as the ground truth image.

\subsubsection{\label{sssec:poca}PoCA reconstruction}
When muons traverse a medium, the Multiple Coulomb Scattering (MCS) occurs. Defining the zenith direction as the Z-axis of a 3D coordinate system, the projected scattering angles onto the XZ or YZ planes can be approximated by a zero-mean Gaussian distribution:
\begin{equation}
    f(\theta)=\frac{1}{\sqrt{2\pi}\,\sigma_{\theta}}\exp\!\left(-\frac{\theta^{2}}{2\sigma^{2}_{\theta}}\right).
    \label{eq:Gaussian}
\end{equation}
The variance of this distribution is determined by the physical properties of the traversed material \cite{Morishima2017}:
\begin{equation}
\begin{gathered}
    \sigma_{\theta}^2
    \approx \left(\frac{13.6\,\mathrm{MeV}}{\beta c p}\right)^2
    \frac{L}{L_{\mathrm{rad}}}.
\end{gathered}
\label{eq:Highland}
\end{equation}
\begin{equation}
    L_{\text{rad}}=\frac{(716.4g cm^{-2})\cdot{A}}{\rho Z(Z+1)\ln\!\left(287/\sqrt{Z}\right)},
    \label{eq:Lrad}
\end{equation}
where $\beta c$ is the muon velocity, $p$ is the muon momentum, $L$ is the material thickness, $L_{rad}$ is the radiation length, $A$ is the mass number of atoms in a material, $Z$ is the atomic number (Z value) of nuclei, and $\rho$ is the density of the material. For the simplification of expression, we define the scattering density:
\begin{equation}
    \lambda=\left(\frac{13.6}{cp_0}\right)^2\frac1{L_{rad}},
\label{scattering density}
\end{equation}
where $p_0$ is the defined reference muon momentum. It is a quantity only related to the radiation length $L_{\text{rad}}$ of a material. These physical relationships constitute the fundamental principle of MST.

The PoCA algorithm is a foundational reconstruction method in MST \cite{Borozdin2003}, which approximates a muon's complex MCS trajectory within a material as a single scattering event. For each muon, the incident and exit trajectories are determined using position data from upper and lower sets of tracking detectors. The PoCA point is then defined as the midpoint of the common perpendicular between these two trajectories. The imaging volume is discretized into a 3D grid of voxels, and the scattering density $\lambda$ is the reconstruction target of the PoCA algorithm. Ignoring momentum and combining Eq. \ref{eq:Highland} and \ref{scattering density}, for each voxel $j$ we can obtain:
\begin{equation}
    \lambda_{j}=\frac{\Theta_{j}}{L_{j}},\quad
    \Theta_{j}=\sum_{i\in E_{j}}\theta_{ij}^{2},\quad
    L_{j}=\sum_{i\in E_{j}} L_{ij}\:,
    \label{eq:lambda}
\end{equation}
where $E_{j}$ is the total count of muon events whose PoCA points fall in $j$, for each valid muon event $i$, the squared scattering angle $\theta^{2}$ is calculated by:
\begin{equation}
    {\theta_i}^2 = \left[\arccos(\frac{\mathbf{v}^{\text{in}}_{i} \cdot \mathbf{v}^{\text{out}}_{i}}{|\mathbf{v}^{\text{in}}_{i}||\mathbf{v}^{\text{out}}_{i}|})\right]^2,
\label{poca weight}
\end{equation}
and $L_{ij}$ is the path length of the muon event $i$ within voxel $j$.

\begin{figure*}[!htb]
\centering
\includegraphics[width=\textwidth]{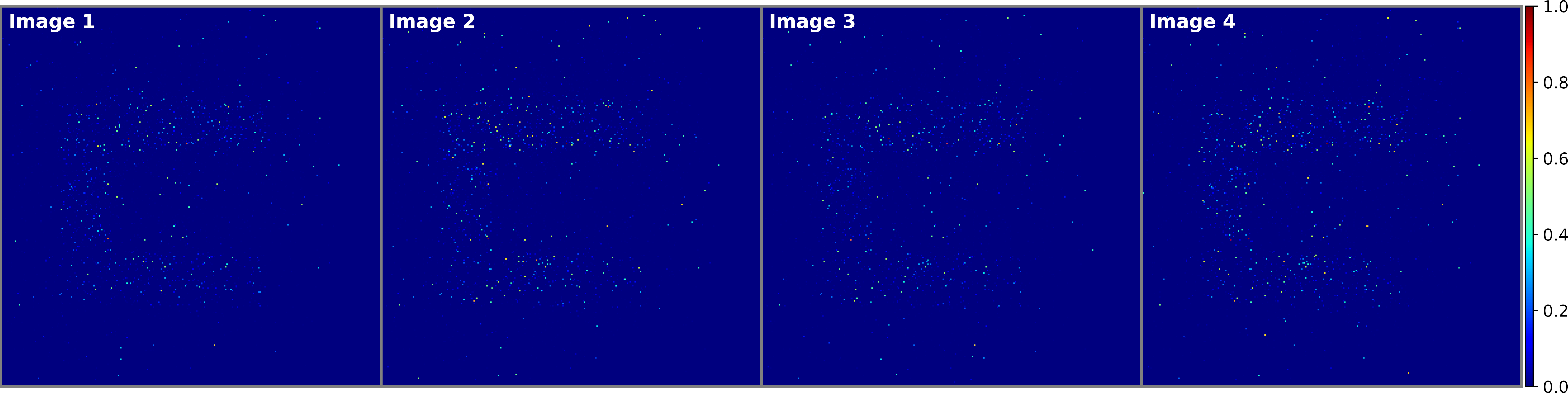}
\caption{Representative PoCA images (\textbf{Image-1} to \textbf{4}) reconstructed from experimental data, the definition of \textbf{Image-1} to \textbf{11} can be found in the Section. \ref{sssec:det}}
\label{fig:pocaexp}
\end{figure*}

\begin{figure*}[!htb]
\centering
\includegraphics[width=\textwidth]{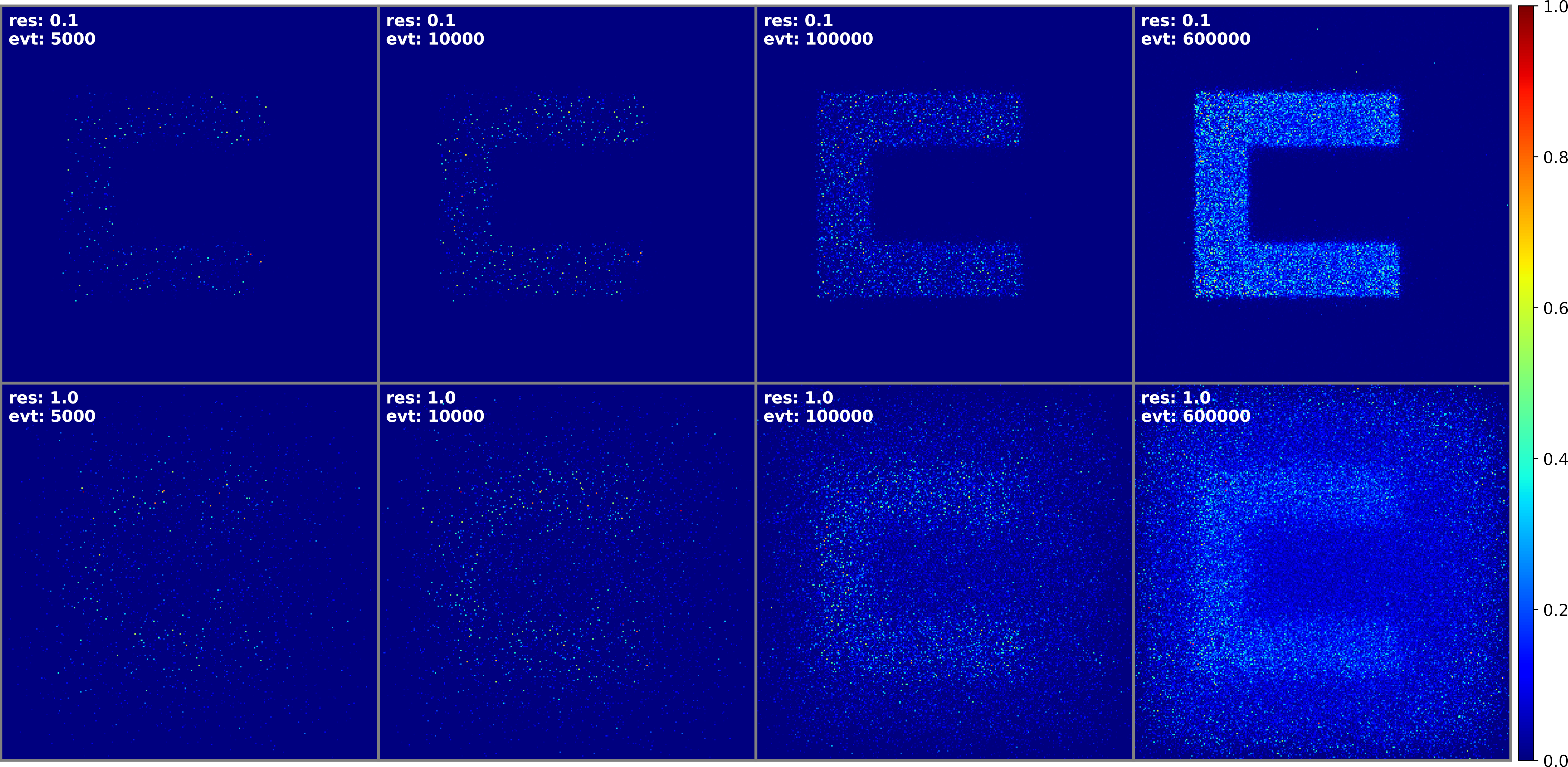}
\caption{Representative PoCA images reconstructed from simulated data, the resolution and the event level of the simulated data corresponding to the image are displayed in the upper left.}
\label{fig:pocasim}
\end{figure*}

More PoCA setting details can be found in our previous work \cite{Yu2024}. We project the 3D voxel grid onto a 2D plane, and the 2D field of view is defined as $\SI{150}{mm}\times\SI{150}{mm}$. This area is discretized into a grid with a pixel size of $\SI{0.5}{mm}\times\SI{0.5}{mm}$, yielding a final 2D image with a resolution of $300\times300$ pixels. Also, to maintain a stable gradient during training and accelerate model convergence, each PoCA image generated subsequently will undergo normalization processing. Fig. \ref{fig:pocaexp} and \ref{fig:pocasim} show some representative PoCA images reconstructed from experimental data and simulated data, respectively. The former is the target to be enhanced, while the latter serves as the dataset for model training. 

\subsection{\label{ssec:stamp}\enquote{Stamping} method}
There has always been a sim-to-real domain gap between the simulated data and the real experimental data that cannot be fully achieved through simulation. This gap is further reflected in the reconstructed images after PoCA. The noise characteristics in real scenes, which blend electronic noise, environmental background, and other systematic artifacts, are far more complex than simple positional resolution. This difference limits the performance of the pure simulation dataset-trained model on images reconstructed with experimental data and may lead to artifacts and structural deficiencies in the enhanced image.

\begin{figure*}[!htb]
\centering
\includegraphics[width=\textwidth]{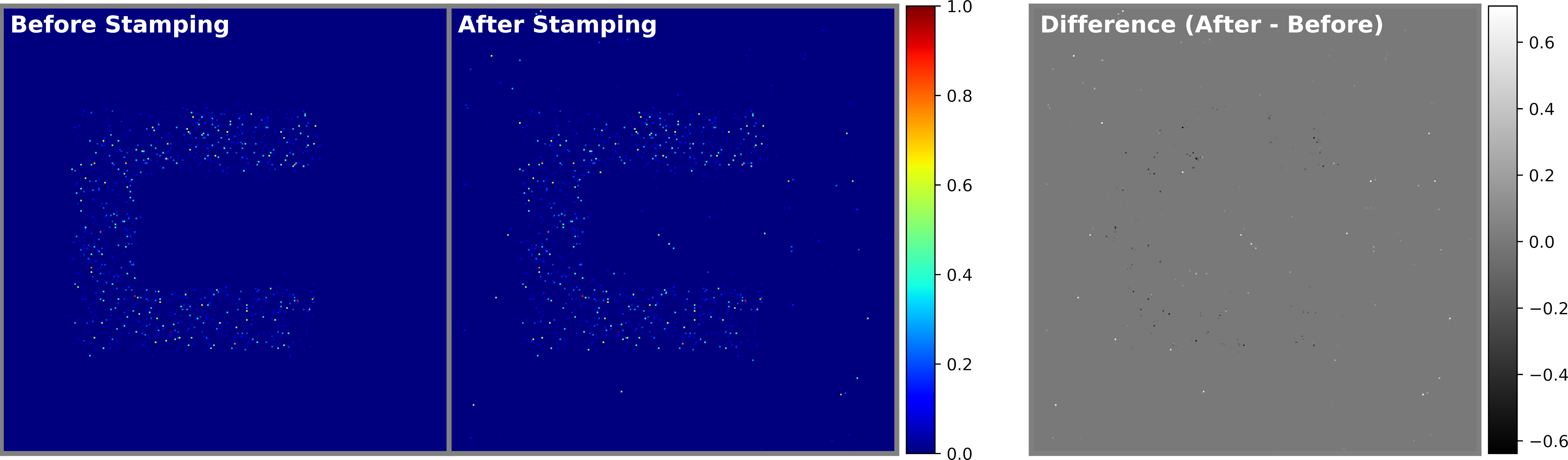}
\caption{The change of a simulated data reconstructed PoCA image before and after \enquote{Stamping} augmentation.}
\label{fig:stamping}
\end{figure*}

To address this key challenge, we propose a novel cross-domain patch injection strategy for training dataset augmentation, termed the \enquote{Stamping} method. The core idea of this method is to inject the unique noise and artifact patterns of the real detector into the pure simulated images, thereby creating a hybrid dataset for domain adaptation training. The specific implementation is shown in Fig. \ref{fig:stamping}. First, we randomly sample $1,000$ small-sized $5\times5$ pixel patches from a real detector image (in this study, we choose \textbf{image 1} as the patch source) with low event level to build a \enquote{Real Style Patch Library}. Subsequently, for each image in the simulated dataset, we randomly select $500$ patches from the \enquote{Library} and directly replace them at random positions in each simulated image. In this way, we constructed augmented simulated images injected with the noise features of real detectors and used them as a cross-domain hybrid dataset. The effectiveness of this \enquote{Stamping} strategy stems from the double randomness it introduces. The randomness of block selection enriches the diversity of training data and prevents the model from overfitting a specific noise pattern. The randomness of the pasting position breaks any false spatial association that may exist between the real noise and the simulated content, forcing the model to learn the statistical characteristics of the noise itself rather than its specific location of occurrence or specific structural information. Meanwhile, by carefully selecting the size and quantity of the blocks, we ensured that the enhanced image still retains its original structural basis and ground truth correspondence. This hybrid dataset enables the training process not only to adapt efficiently to the real domain but also to maintain stability.

\subsection{\label{ssec:quality}Image quality assessment}
IQA is an objective approach to evaluate image quality. To quantitatively and comprehensively evaluate the performance of the image enhancement framework, we adopted four complementary image quality assessment metrics. These metrics measure the consistency between the enhanced image and the ground truth image from different dimensions, including pixel-level fidelity, structural similarity, shape matching degree, and human perception similarity. Use $X$ to represent the enhanced image output by the model, and $Y$ to represent the enhancement objective, that is, the ground truth image. We introduce the following evaluation metrics:

\textbf{Peak Signal-to-Noise Ratio (PSNR): }PSNR \cite{5596999} is one of the most commonly used and fundamental indicators in image quality assessment. It is calculated based on the Mean Squared Error (MSE) between the enhanced image and the ground truth image, reflecting the magnitude of the error at the pixel level:
\begin{equation}
    \text{PSNR}(X, Y) = 10 \cdot \log_{10} \left( \frac{MAX_X^2}{\text{MSE}(X, Y)} \right),
    \label{psnr}
\end{equation}
where
\begin{equation}
    \text{MSE}(X, Y) = \frac{1}{N} \sum_{i=1}^{N} (X_i - Y_i)^2,
    \label{mse}
\end{equation}
where $i$ represents the pixel number. The higher the value, the smaller the image distortion and the better the quality. In our study, we import \textit{peak\_signal\_noise\_ratio} from \textit{skimage.metrics}. However, PSNR only focuses on the absolute differences in pixel values, and it is overly sensitive to brightness.

\textbf{Intersection over Union (IoU): }In our task, merely restoring the brightness and structure of the pixels is not enough. Ensuring the geometric integrity of the shape is also of crucial importance. For this purpose, we have introduced the IoU metric \cite{Yu_2016}. When calculating, we first set a $0.1$ threshold to convert both the ground truth image and the image to be evaluated into binarized masks. IoU measures the degree of overlap between the predicted shape and the actual shape by calculating the ratio of the intersection area to the union area of these two masks:
\begin{equation}
    \text{IoU}(x, y) = \frac{|X \cap Y|}{|X \cup Y|},
    \label{iou}
\end{equation}
the higher the value, the more consistent the predicted structural shape is with the actual structure, and the better the conformal property. However, the evaluation result is highly dependent on the selection of the threshold.

\textbf{Structural Similarity Index Measure (SSIM): }SSIM \cite{ZhouWang2004, 5596999} models the similarity of local regions in an image from luminance, contrast, and structure. It is more in line with the perception characteristics of the human visual system than PSNR:
\begin{equation}
    \text{SSIM}(x, y) = 
    \frac{(2\mu_x \mu_y + C_1)(2\sigma_{xy} + C_2)}
    {(\mu_x^2 + \mu_y^2 + C_1)(\sigma_x^2 + \sigma_y^2 + C_2)},
    \label{ssim}
\end{equation}
where $\mu$ is the \textit{local} mean value, $\sigma$ is the \textit{local} variance or covariance, $C_1$ and $C_2$ are constants related to pixel brightness to prevent the denominator from taking zero. Taking the average of the SSIM values of all local areas to obtain the final SSIM value of the entire image:
\begin{equation}
    \text{SSIM}(X, Y) = \frac{1}{M} \sum_{j=1}^{M} \text{SSIM}(x_j, y_j)
    \label{ssim}
\end{equation}
The value range of SSIM is from $-1$ to $1$. The higher the value, the more similar the structure of the two images is, and the better the quality. In our study, we import \textit{structural\_similarity} from \textit{skimage.metrics} and all parameter settings are default values. However, it is easily deceived by large areas of simple backgrounds, resulting in a relatively high SSIM score even if the similarity of the structure to be evaluated is very low. Therefore, attention should be focused on the changes in SSIM rather than its value.

\textbf{Learned Perceptual Image Patch Similarity (LPIPS): }To introduce a deeper and more human-like assessment, we adopted the LPIPS \cite{Zhang2018}. Unlike traditional metrics that rely on fixed mathematical formulas, LPIPS utilizes a deep neural network (VGG) pre-trained on large image datasets (ImageNet) as a feature extractor. It measures the perceptual similarity of two images by comparing the differences in their feature maps at different depths of the network, from low-level textures to high-level semantic features. LPIPS can effectively capture artifacts, blurriness, or unrealistic textures that are easily overlooked by traditional metrics. It should be particularly noted that LPIPS is a distance or loss indicator. Therefore, the lower its value, the more similar the two images are to human perception and the better their quality.

In summary, by combining the use of PSNR, IoU, SSIM, and LPIPS, we have constructed a multi-dimensional and robust evaluation system that can conduct a comprehensive and in-depth analysis of the optimization performance of our model.

\section{\label{sec:result}Result and discussion}

\subsection{\label{ssec:sampling}Training dataset setup}
\begin{figure}[!htb]
\includegraphics[width=\linewidth]{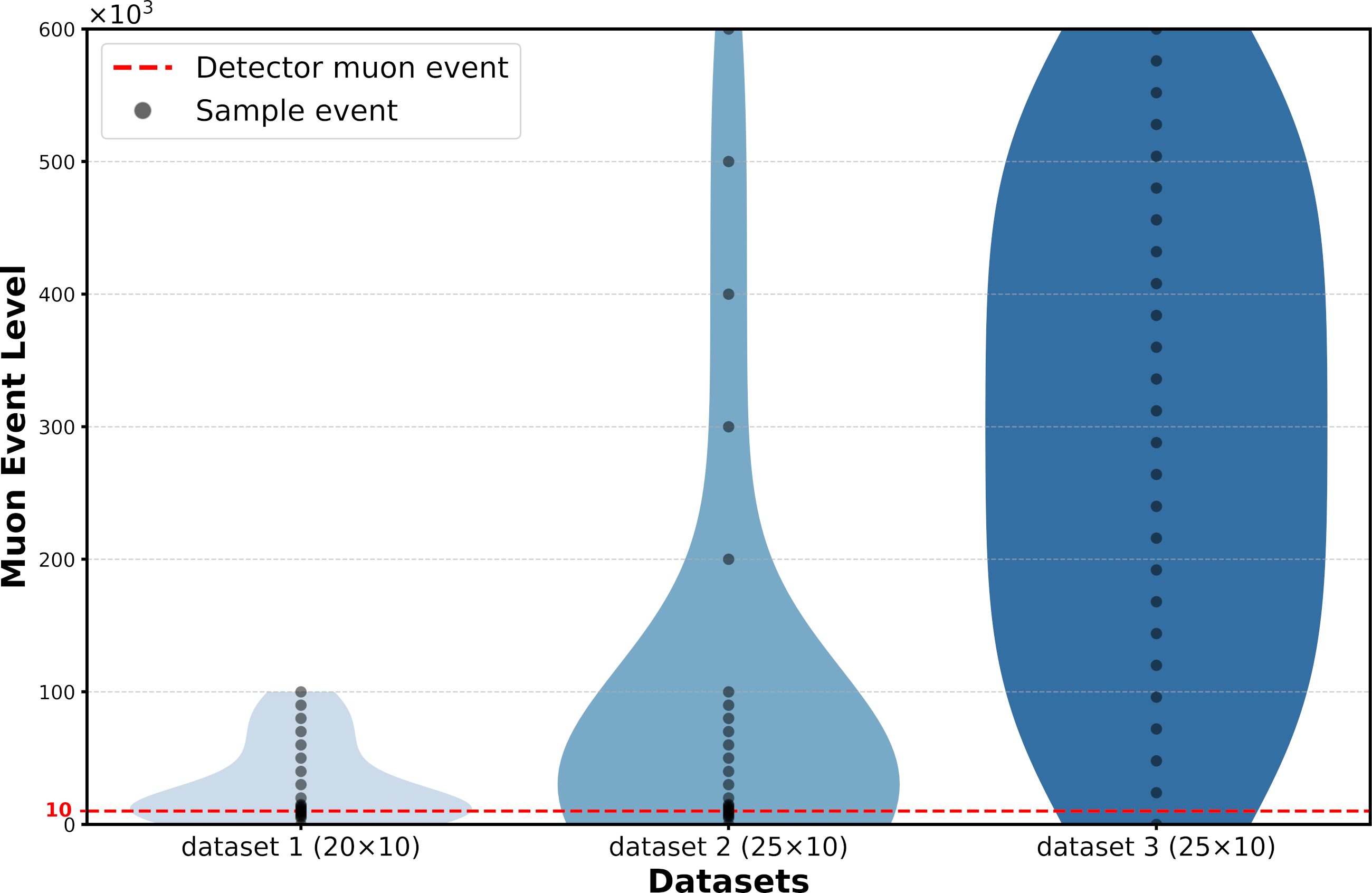}
\caption{Violin plot of the densities of the three dataset setting strategies in different event levels. The final sample number in dataset is $\times10$ based on 10 simulated position resolutions.}
\label{fig:datasets}
\end{figure}

The efficacy of a hybrid dataset generated via the same \enquote{Stamping} method augmentation is also dependent on the distribution of the underlying simulation data. We therefore investigated three distinct setting strategies for constructing the simulation dataset, as illustrated in Fig. \ref{fig:datasets}.

\begin{enumerate}
\itemsep=0pt
\item {\textbf{Dataset-1}: narrow, low-event-count setting.} An intuitive approach where the event counts of simulated images are narrowly concentrated around the level of the experimental data.
\item {\textbf{Dataset-2}: focused, wide-range setting.} Our proposed optimal strategy, which combines a broad range with a focused core. The sampling range is wide, spanning from the experimental event-count level to the ground truth to provide a global context. While the sampling density is also biased toward the low-event-count region most relevant to our enhancement task.
\item {\textbf{Dataset-3}: uniform full-range setting.} This strategy involves uniformly sampling images across the entire spectrum, from low-event-count levels up to the event count of ground truth.
\end{enumerate} 

\begin{figure*}[!htb]
\centering
\includegraphics[width=\textwidth]{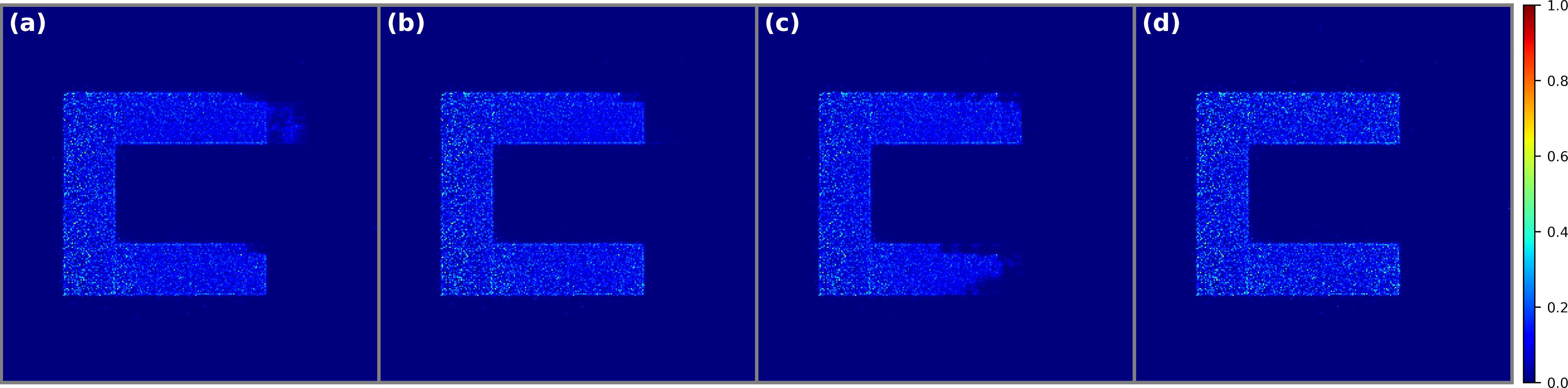}
\caption{The enhancement results of \textbf{Image-2}  (defined in Section. \ref{sssec:sim}) based on models trained with different hybrid datasets and the ground truth image. (a) \textbf{Dataset-1}-based; (b) \textbf{Dataset-2}-based; (c) \textbf{Dataset-3}-based; (d) ground truth image.}
\label{fig:pretrain}
\end{figure*}

We evaluated these three strategies by training a separate model on each dataset after the same \enquote{Stamping} augmentation and comparing their enhancement performance on \textbf{Image-2}, with the results visualized in Fig. \ref{fig:pretrain}. The model trained with \textbf{Dataset-1} yielded suboptimal results (Fig. \ref{fig:pretrain}(a)). Because it was exposed only to the images with low-event-count, the model failed to learn the fundamental statistical distinction between signal and noise, causing it to amplify dense noise regions into severe structural artifacts (hallucinations). \textbf{Dataset-3} also proved ineffective (Fig. \ref{fig:pretrain}(c)). The uniform sampling led to insufficient training exposure in the critical low-event-count regime, resulting in a model with a limited ability to distinguish weak signals from background noise, leading to incomplete structures and loss of detail. In contrast, the model trained with our proposed \textbf{Dataset-2} demonstrated significantly superior enhancement performance (Fig. \ref{fig:pretrain}(b)), effectively suppressing noise while preserving structural integrity.

\begin{table*}[!htb]
	\centering
	\caption{Comparison of the impact of different loss terms in the training stage.}
	\label{tab:loss}
	
	\renewcommand{\arraystretch}{1.2} % 调整行间距
	\begin{tabular}{>{\raggedright\arraybackslash}p{1.9cm} 
			>{\centering\arraybackslash}p{1.3cm} 
			>{\centering\arraybackslash}p{1.3cm} 
			>{\centering\arraybackslash}p{1.3cm}
            >{\centering\arraybackslash}p{1.3cm}
            >{\centering\arraybackslash}p{2.3cm}
            >{\centering\arraybackslash}p{2.3cm}
            >{\centering\arraybackslash}p{2.3cm}}
		
		\toprule[1pt]
		\multirow{2}{*}[-0.4ex]{\textbf{Loss term}} & \multicolumn{4}{c}{\textbf{Metrics}} & \multirow{2}{*}{\shortstack{\textbf{Total time} \\ \textbf{(300 epochs)}}} & \multirow{2}{*}{\shortstack{\textbf{Peak VRAM} \\ \textbf{Usage (GB)}}} & \multirow{2}{*}{\shortstack{\textbf{Peak RAM} \\ \textbf{Usage (GB)}}} \\
		
		\cmidrule(l){2-5}
		& PSNR$\uparrow$ & IoU$\uparrow$ & SSIM$\uparrow$ & LPIPS$\downarrow$ \\
		
		\midrule
		SSIM & 33.21 & 0.7476 & 0.9507 & 0.0660 & \textasciitilde61.8min & \textasciitilde2.82 & \textasciitilde10.50 \\
		GSSIM & 35.45 & 0.7877 & 0.9627 & 0.0634 & \textasciitilde62.7min & \textasciitilde2.82 & \textasciitilde10.53 \\
		LPIPS & 36.72 & 0.8209 & 0.9729 & 0.0241 & \textasciitilde104.4min & \textasciitilde3.80 & \textasciitilde10.72 \\
		
		\bottomrule[1pt]
	\end{tabular}
\end{table*}
\subsection{\label{ssec:loss}Training with different loss}
A controlled experiment is conducted to identify the optimal loss function for our study (Table. \ref{tab:loss}). The LPIPS component of our joint loss function was systematically replaced with Global-SSIM (GSSIM) or the SSIM. The calculation of GSSIM is to replace the local patch area $x$ and $y$ by the global image $X$ and $Y$. Our findings revealed that employing both the GSSIM-based loss and the SSIM-based loss results in suboptimal model performance. Conversely, the LPIPS-based loss achieved the best performance in all IQA scores. This indicates that compared to the statistical-based SSIM, LPIPS has better image perception capabilities with comparisons of multi-layer scales and semantics. Therefore, the joint loss function that incorporates the LPIPS loss term is used as the final training loss. However, it is worth noting that LPIPS introduced a larger computational overhead. Considering the trade-off between performance and efficiency, the joint loss based on SSIM can be an alternative when computing resources and time are tight. In addition, due to the sliding window mechanism that comes with SSIM calculation, when the thickness or material density of the image object is uneven, the joint loss function based on the SSIM term may lead to better training performance.

\begin{figure*}[!htb]
\centering
\includegraphics[width=\textwidth]{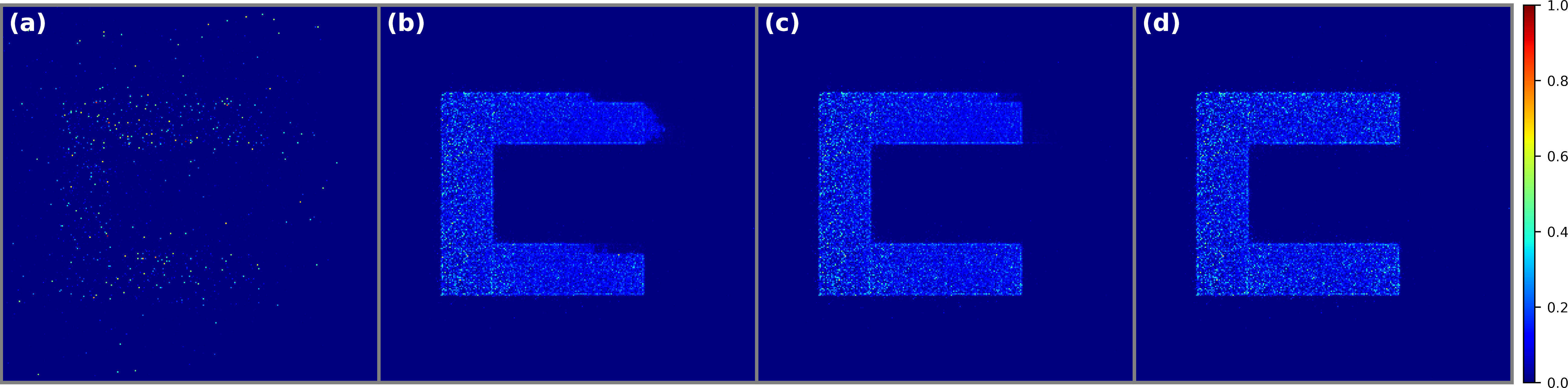}
\caption{The comparison between (a) the original PoCA reconstructed \textbf{Image-2}; (b) enhancement output of model trained with simulation dataset; (c) enhancement output of model trained with hybrid dataset; (d) ground truth image.}
\label{fig:ablation}
\end{figure*}

\subsection{\label{ssec:ablation}Ablation and generalization study}
To validate the efficacy of our proposed \enquote{Stamping} method in improving model performance, i.e., a better enhancement effect on experimental data reconstructed PoCA images, we conducted an ablation study. Specifically, we trained the model using the pure simulation dataset without \enquote{Stamping}. As shown in Fig. \ref{fig:ablation}, compared to the model trained on the hybrid dataset, the image enhanced by the pure simulation dataset-based model has lower scores on all four IQA metrics (Table. \ref{tab:result}). More importantly, the enhanced image had more severe structural defects and background artifacts. These findings validate the effectiveness of the \enquote{Stamping} based hybrid dataset training in improving the model's enhancement performance on real experimental data reconstructed images.

\begin{figure*}[!htb]
\centering
\includegraphics[width=\textwidth]{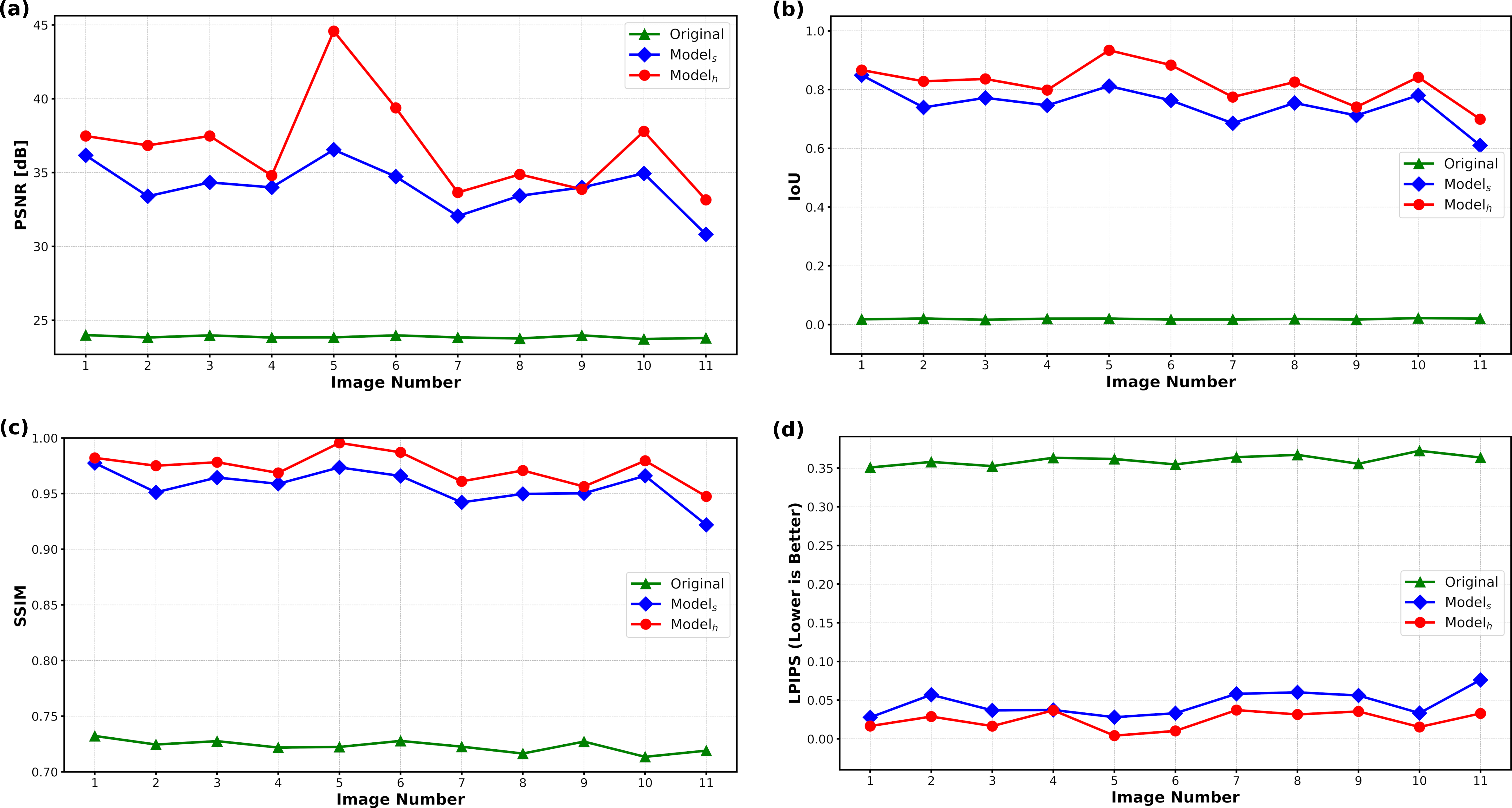}
\caption{The comparison (from \textbf{Image-1} to \textbf{11}) of four metrics between the original images, images enhanced by the model trained on the simulation dataset (Model\textsubscript{s}), images enhanced by the model trained on the hybrid dataset (Model\textsubscript{h}): (a) PSNR; (b)IoU; (c) SSIM; (d) LPIPS.}
\label{fig:comparison}
\end{figure*}

Furthermore, to assess the generalization capability of our training framework, that is, a model trained on data augmented with just one experimental data reconstructed image can also improve the enhancement results for other images from the same detector under the same event level. We used the trained model to enhance the total 11 images (\textbf{Image-1} to \textbf{11}). The results show that, except for the PSNR of \textbf{Image-9}, the enhancement results of the hybrid dataset-based model for the 11 images were comprehensively improved on the other metrics compared to the simulation dataset-based model (Fig. \ref{fig:comparison}). The reason for the slight decrease in PSNR on individual images is that it is heavily affected by the overall brightness of the image.

\begin{table}[h]
	\centering
	\caption{The metric scores of \textbf{Image-2} before and after enhancement, where Original means the experimental data reconstructed PoCA image before enhancement, Model\textsubscript{s} means the enhancement output of the model trained on the simulation dataset, and Model\textsubscript{h} means the enhancement output of the model trained on the hybrid dataset.}
	\label{tab:result}
	
	\renewcommand{\arraystretch}{1.2} % 调整行间距
	\begin{tabular}{>{\raggedright\arraybackslash}p{1.6cm} 
			>{\centering\arraybackslash}p{1.2cm} 
			>{\centering\arraybackslash}p{1.2cm} 
			>{\centering\arraybackslash}p{1.2cm}
            >{\centering\arraybackslash}p{1.2cm}}
		
		\toprule[1pt]
		\multirow{2}{*}[-0.4ex]{\makecell[l]{\textbf{Enhencement}\\\textbf{stage}}} & \multicolumn{4}{c}{\textbf{Metrics}}\\
		
		\cmidrule(l){2-5}
		& PSNR$\uparrow$ & IoU$\uparrow$ & SSIM$\uparrow$ & LPIPS$\downarrow$ \\
            
            \midrule
		Original & 23.83 & 0.0206 & 0.7245 & 0.3581 \\
		Model\textsubscript{s} & 33.40 & 0.7393 & 0.9512 & 0.0570 \\
		Model\textsubscript{h} & \textbf{36.84} & \textbf{0.8281} & \textbf{0.9750} & \textbf{0.0288} \\
        
		\bottomrule[1pt]
	\end{tabular}
\end{table}

\begin{table}[h]
	\centering
	\caption{The average metric scores of \textbf{Image-1} to \textbf{11} before and after enhancement. The meaning of the table header is the same as Table. \ref{tab:result}.}
	\label{tab:average_result}
	
	\renewcommand{\arraystretch}{1.2}
	
	\begin{tabular}{>{\raggedright\arraybackslash}p{1.6cm} 
			>{\centering\arraybackslash}p{1.2cm} 
			>{\centering\arraybackslash}p{1.2cm} 
			>{\centering\arraybackslash}p{1.2cm}
            >{\centering\arraybackslash}p{1.2cm}}
		
		\toprule[1pt]
		\multirow{2}{*}[-0.4ex]{\makecell[l]{\textbf{Enhencement}\\\textbf{stage}}} & \multicolumn{4}{c}{\textbf{Metrics}}\\
		
		\cmidrule(l){2-5}
		& PSNR$\uparrow$ & IoU$\uparrow$ & SSIM$\uparrow$ & LPIPS$\downarrow$ \\
		
		\midrule
		Original & 23.87 & 0.0189 & 0.7232 & 0.3604 \\
		Model\textsubscript{s} & 33.71 & 0.7450 & 0.9526 & 0.0512 \\
		Model\textsubscript{h} & \textbf{36.17} & \textbf{0.8099} & \textbf{0.9699} & \textbf{0.0270} \\
		
		\bottomrule[1pt]
	\end{tabular}
\end{table}

\section{\label{sec:conclusion}Conclusion}
This study proposes a U-Net-based image enhancement framework aimed at significantly improving the image quality of short-time MST. By learning the mapping from low-event-count image distributions to high-event-count image distributions, the framework effectively reconstructs noisy, low-quality inputs into clear images, with visual quality and quantitative metrics comparable to those obtained from long-time MST.

In this framework, a tailored U-Net architecture was specifically designed for the MST image enhancement task. Given the extreme scarcity of experimental MST data, this study fully leverages MST simulation tools to construct a large-scale and diverse PoCA image dataset, providing high-quality and reliable training samples for the U-Net model. In addition, a systematic comparison of various loss functions was conducted to achieve optimal enhancement performance. Furthermore, a novel data augmentation strategy was introduced, which further improves the model's ability to preserve structural information and suppress background noise.

To comprehensively evaluate the effectiveness of the proposed method, a series of widely recognized image quality assessment metrics were employed, including Peak Signal-to-Noise Ratio (PSNR), Intersection over Union (IoU), Structural Similarity Index (SSIM), and Learned Perceptual Image Patch Similarity (LPIPS), for quantitative comparison of images before and after enhancement. The results demonstrate that the trained model exhibits excellent performance on experimental MST images, showing outstanding generalization capability and enhancement quality. This framework substantially mitigates the key bottlenecks of MST related to acquisition time and detector resolution, enabling better adaptation to time-sensitive application scenarios while significantly reducing reliance on scarce experimental MST data.

Future work will extend the framework to more complex imaging structures and explore unsupervised or self-supervised learning paradigms based on advanced models, with the expectation of further enhancing the practicality and accessibility of muon tomography as a powerful and broadly applicable imaging tool.
Future work will extend the framework to more complex imaging structures and explore unsupervised or self-supervised learning paradigms based on advanced models, with the expectation of further enhancing the practicality and accessibility of muon tomography as a powerful and broadly applicable imaging tool.Future work will extend the framework to more complex imaging structures and explore unsupervised or self-supervised learning paradigms based on advanced models, with the expectation of further enhancing the practicality and accessibility of muon tomography as a powerful and broadly applicable imaging tool.

\printcredits

\section*{\label{sec:declaration}Declaration of competing interest}
The authors declare that they have no known competing financial interests or personal relationships that could have appeared to influence the work reported in this paper.

\section*{\label{sec:data}Data availability}
Data will be made available on request.

\section*{\label{sec:acknowledgments}Acknowledgments}
This work is supported by the Research Program of State Key Laboratory of Heavy Ion Science and Technology, Institute of Modern Physics, Chinese Academy of Sciences, under Grant No. HIST2025CS06, the National Natural Science Foundation of China (Grant No. 12105327, 12475106), the Guangdong Basic and Applied Basic Research Foundation (Grant No. 2023B1515120067), and the Fundamental Research Funds for the Central Universities under Grant No. JZ2025HGTG0252. This work is partly supported by High Intensity heavy-ion Accelerator Facility (HIAF) project approved by the National Development and Reform Commission of China.

%\newpage

\bibliographystyle{model1-num-names}

\bibliography{refs}

\end{document}